\documentclass[12pt]{article}
\usepackage[dvips]{graphicx}

\usepackage{graphicx}
\usepackage{dcolumn}
\usepackage{amsmath,amsfonts}
\usepackage{bm}

\usepackage{epsfig,psfrag,color}
\usepackage{subfig}
\usepackage{amsmath}
\usepackage{amssymb}
\usepackage{amsfonts}
\usepackage{latexsym}
\usepackage{times}
\usepackage{subfig}
\usepackage{arydshln}
\usepackage{lineno}
\usepackage{setspace}
\usepackage{caption}

\usepackage[text={6in,8.75in},centering,dvipdfm]{geometry}

\newcommand{\real}[1]{{\mathbb R}^{#1}}
\newcommand{\sreal}[1]{{\mathbb S}^{#1}}

\newcommand{\spn}{\mathrm{span}}

\newcommand{\tr}{\mathrm{\,tr}}
\newcommand{\vecc}{\mathrm{\,vec}}

\newcommand{\I}{\mathbf I}

\newcommand{\start}{\mathrm{start}}
\newcommand{\oneD}{\mathrm{1D}}

\newcommand{\X}{\mathbf X}

\newcommand{\Y}{{\mathbf Y}}

\newcommand{\abf}{\mathbf a}

\newcommand{\g}{{\mathbf g}}
\newcommand{\gt}{\tilde{\mathbf g}}
\newcommand{\Obf}{\mathbf O}
\newcommand{\ubf}{\mathbf u}
\newcommand{\ubfhat}{\widehat{\ubf}}

\newcommand{\Sbf}{\mathbf{S}}

\makeatletter
\renewcommand*{\@seccntformat}[1]{%
   \csname the#1\endcsname.\quad}
\makeatother

\newcommand{\G}{\mathbf{G}}

\newcommand{\Pbf}{{\mathbf P}}

\newcommand{\V}{{\mathbf V}}
\newcommand{\Vhat}{\widehat \V}
\newcommand{\bv}{{\mathbf v}}

\newcommand{\U}{{\mathbf U}}
\newcommand{\Uhat}{\widehat \U}

\newcommand{\W}{{\mathbf W}}
\newcommand{\What}{\widehat{\W}}

\newcommand{\M}{{\mathbf M}}

\newcommand{\s}{{\mathbf s}}

\newcommand{\A}{{\mathbf A}}
\newcommand{\Ahat}{\widehat{\mathbf A}}

\newcommand{\B}{{\mathbf B}}
\newcommand{\Q}{{\mathbf Q}}
\newcommand{\C}{{\mathbf C}}

\newcommand{\greekbold}[1]{\mbox{\boldmath $#1$}}
\newcommand{\alphabf}{\greekbold{\alpha}}

\newcommand{\etabf}{\greekbold{\eta}}

\newcommand{\betabf}{\greekbold{\beta}}

\newcommand{\betabfhat}{\widehat{\greekbold{\beta}}}

\newcommand{\Gammabf}{\greekbold{\Gamma}}

\newcommand{\Gammabfhat}{\widehat{\greekbold{\Gamma}}}

\newcommand{\Omegabf}{\greekbold{\Omega}}

\newcommand{\phibf}{\greekbold{\phi}}
\newcommand{\phibft}{\widetilde{\greekbold{\phi}}}
\newcommand{\phibfs}{\greekbold{\scriptstyle \phi}}

\newcommand{\psibf}{\greekbold{\psi}}

\newcommand{\Sigmabf}{\greekbold{\Sigma}}

\newcommand{\Sigmabfs}{{\greekbold{\scriptstyle \Sigma}}}

\newcommand{\thetabf}{\greekbold{\theta}}
\newcommand{\thetabft}{\widetilde {\greekbold{\theta}}}
\newcommand{\thetabfs}{\greekbold{\scriptstyle \theta}}
\newcommand{\Thetabf}{\greekbold{\Theta}}

\newcommand{\varepsilonbf}{\greekbold{\varepsilon}}

\newcommand{\spc}{{\mathcal S}}

\newcommand{\Mhat}{\widehat{{\mathbf M}}}

\newcommand{\aspc}{\mathcal A}
\newcommand{\bspc}{\mathcal B}
\newcommand{\rspc}{\mathcal R}
\newcommand{\espc}{{\cal E}}
\newcommand{\espchat}{\widehat{\cal E}}

\newcommand{\gspc}{{\mathcal G}}

\newcommand{\uspc}{{\mathcal U}}
\newcommand{\vspc}{{\mathcal V}}

\newcommand{\semu}{{\cal E}_{\M}(\uspc)}
\newcommand{\semuhat}{\widehat{{\cal E}}_{\M}(\uspc)}

\newcommand{\seb}{{\cal E}_{\Sigmabfs}(\bspc)}

\newcommand{\isebjhat}[1]{\widehat{{\cal IE}}_{\Sigmabfs}(\bspc)}

\newcommand{\noboxfig}[2]{\includegraphics[width=#2in]{#1.eps}}
\newcommand{\boxfig}[2]{\fbox{\noboxfig{#1}{#2}}}

\newcommand{\smallfig}{2.38}

\newcommand{\Dfig}[2]{\noboxfig{#1}{#2}}
\newcommand{\Cfig}[2]{\centerline{\psfig{figure=#1.eps,width=#2in}}}

  {\addtolength{\leftskip}{#1}\addtolength{\rightskip}{#2}}{\par}

\newtheorem{theorem}{Theorem}[section]

\newtheorem{proposition}[theorem]{Proposition}

\begin{document}
\title{{Algorithms for Envelope Estimation II}}

\author{R. Dennis Cook%
\thanks{{\small R. Dennis Cook is Professor, School of Statistics, University
of Minnesota, Minneapolis, MN 55455 (E-mail: dennis@stat.umn.edu).}%
}, \, Liliana Forzani%
\thanks{{\small Liliana Forzani is Professor, Facultad de Ingenier\'ia Qu\'imica, Universidad Nacional del Litoral and Instituto Matem\'atica Aplicada Litoral, CONICET-UNL, Santa Fe, Argentina (Email: liliana.forzani@gmail.com). }%
} 
\ and Zhihua Su\thanks{{\small Zhihua Su is Assistant Professor, Department of Statistics, University
of Florida, Gainsville, FL 32611-8545  (E-mail: zhihuasu@stat.ufl.edu).}}}
\maketitle

\begin{abstract}
We propose a new algorithm for envelope estimation, along with a new $\sqrt{n}$-consistent method for computing starting values.  The new algorithm, which does not require optimization over a Grassmannian, is shown by simulation to be much faster and typically more accurate that the best existing algorithm proposed by Cook and Zhang (2015c).  

\end{abstract}
\section{Introduction}\label{sec:intro}

The  goal of envelope  methods is to increase efficiency in multivariate parameter estimation and prediction by exploiting variation in the data  that is effectively immaterial to the goals of the analysis.  Envelopes achieve efficiency gains by basing estimation on the variation that is material to those goals, while simultaneously excluding that which is immaterial.  It now seems evident that immaterial variation is often present in multivariate analyses and that the estimative improvement afforded by envelopes  can be quite substantial when the immaterial variation is large, sometimes equivalent to taking thousands of additional observations.

  Algorithms  for envelope estimation require optimization of a non-convex objective function over a Grassmannian, which can be quite slow in all but small or modest sized problems, possibly taking hours or even days to complete an analysis of a sizable problem.  Local optima are another complication that may increase the difficulty of the computations and the analysis generally. Until recently, envelope methods were available only in Matlab, as these computing issues hindered  implementation in R. 

In this article we propose new easily computed $\sqrt{n}$-consistent starting values and a novel non-Grassmann algorithm for optimization of the most common envelope objective function.  These computing tools are much faster than current algorithms in sizable problems and can be implemented straightforwardly in R.  The new starting values have proven quite effective and can be used as fast standalone estimators in exploratory analyses.

 In the remainder of this introduction we review envelopes and describe the computing issues in more detail. We  let $\Pbf_{(\cdot)}$ denote a projection with $\Q_{(\cdot)} = \I - \Pbf_{(\cdot)}$,  let $\real{r\times c}$ be the set of all real $r\times c$ matrices, and let $\sreal{k\times k}$ be the set of all real and symmetric
$k\times k$ matrices.
If $\M\in\real{r\times c}$, then $\spn(\M)\subseteq\real{r}$
is the subspace spanned by columns of $\M$.  $\vecc$ is the vectorization operator that stacks the columns of a matrix.  A subspace $\rspc\subseteq\real{p}$
is said to be a reducing subspace of $\M\in\real{p\times p}$ if
$\rspc$ decomposes $\M$ as $\M=\Pbf_{\rspc}\M\Pbf_{\rspc}+\Q_{\rspc}\M\Q_{\rspc}$.
If $\rspc$ is a reducing subspace of $\M$, we say that $\rspc$
reduces $\M$.

\subsection{Review of envelopes}\label{sec:review}

Envelopes were originally proposed and developed by Cook,  Li and Chiaromonte (2007, 2010) in the context of  multivariate linear regression, 
\begin{equation}\label{mlm}
\Y_{i} = \alphabf + \betabf \X_{i} + \varepsilonbf_{i},\ i=1,\dots,n,
\end{equation}
where $\varepsilonbf_i\in\real{r}$ is a normal error vector with mean 0,  variance $\Sigmabf>0$ and is independent of $\X$, $\alphabf \in \real{r}$ and $\betabf\in\real{r\times p}$ is the regression coefficient matrix in which we are primarily interested. Immaterial variation can occur in $\Y$ or $\X$ or both.  Cook et al. (2010) operationalized the idea of immaterial variation in the response vector by asking if there are linear combinations of $\Y$ whose distribution is invariant to changes in $\X$.  Specifically, let $\Pbf_{\espc}\Y$ denote the projection onto a subspace $\espc\subseteq\real{r}$ with the properties  (1) the distribution of $\Q_{\espc}\Y \mid \X$ does not depend on the value of  the non-stochastic predictor $\X$ and (2) $\Pbf_{\espc}\Y$ is independent of $\Q_{\espc}\Y$ given $\X$.  These conditions imply that the distribution of $\Q_{\espc}\Y$ is not affected by $\X$ marginally or through an association with $\Pbf_{\espc}\Y$.
Consequently, changes in the predictor affect the distribution of $\Y$ only via $\Pbf_{\espc}\Y$ and so we refer to  $\Pbf_{\espc}\Y$ informally as the material part of $\Y$ and to $\Q_{\espc}\Y$ as the  immaterial part of $\Y$.  

Conditions (1) and (2) hold if and only if (a) $\bspc := \spn(\betabf) \subseteq \espc$ (so $\espc$ {\em envelopes} $\bspc$) and (b) $\espc$ reduces $\Sigmabf$.  The $\Sigmabf$-envelope of $\bspc$, denoted $\seb$, is  defined formally as the intersection of all reducing subspaces  of $\Sigmabf$ that contain $\bspc$.  Let $u= \dim(\espc_{\Sigmabfs}(\betabf))$ and  let $(\Gammabf, \Gammabf_{0}) \in \real{r\times r}$ be an orthogonal matrix with $\Gammabf \in \real{r \times u}$ and $\spn(\Gammabf) = \espc_{\Sigmabfs}(\bspc)$.    
This leads directly to the envelope version of model (\ref{mlm}),
\begin{equation}\label{envmodel}
\Y_{i} = \alphabf + \Gammabf \etabf \X_{i} + \varepsilonbf_{i}, \mathrm{\;with\;}  \Sigmabf = \Gammabf \Omegabf \Gammabf^{T} + \Gammabf_{0} \Omegabf_{0} \Gammabf_{0}^{T}, \;i=1,\ldots,n,
\end{equation}
where $\betabf=\Gammabf\etabf$, $\etabf \in \real{u \times p}$ gives the coordinates of $\betabf$ relative to basis $\Gammabf$, and   $\Omegabf \in \sreal{u \times u}$ and $\Omegabf_{0} \in \sreal{(r-u)\times(r-u)}$ are positive definite matrices. While  $\etabf$, $\Omegabf$ and $\Omegabf_{0}$  depend on the basis $\Gammabf$ selected to represent $\espc_{\Sigmabfs}(\betabf)$, the parameters of interest $\betabf$ and $\Sigmabf$ depend only on $\espc_{\Sigmabfs}(\betabf)$ and not on the basis.  All parameters in (\ref{envmodel}) can be estimated by maximizing its likelihood with the envelope dimension $u$ determined by using standard methods like likelihood ratio testing, information criteria, cross-validation or a hold-out sample, as described by Cook et al. (2010). The envelope estimator $\betabfhat$ of $\betabf$ is just the projection of the maximum likelihood estimator $\B$ onto the estimated envelope, $\betabfhat = \Pbf_{\espchat}\B$, and $\sqrt{n}(\vecc(\betabfhat) - \vecc(\betabf))$ is asymptotically normal with mean 0 and covariance matrix given by Cook et al. (2010), where  $u$ is assumed to be known.  An introductory example of response envelopes is available in Cook and Zhang (2015a).

Similar reasoning leads to partial envelopes for use when only selected columns of $\betabf$ are of interest (Su and Cook, 2011), to predictor envelopes allowing for immaterial variation in $\X$ (Cook, Helland and Su, 2013), to predictor-response envelopes allowing simultaneously for immaterial variation in $\X$ and $\Y$ (Cook and Zhang, 2015b) and to heteroscedastic envelopes for comparing the means of multivariate populations with unequal covariance matrices (Su and Cook, 2013).

Cook and Zhang (2015a) extended envelopes beyond multivariate linear models by proposing the following estimative construct for vector-valued parameters.  Let $\thetabft$ denote an estimator of a parameter
vector $\thetabf\in\Thetabf \subseteq \real{m}$
based on a sample of size $n$ and assume, as is often the case, that $\sqrt{n}(\thetabft-\thetabf)$
converges in distribution to a normal random vector with mean 0 and
covariance matrix $\V(\thetabf)>0$ as $n\rightarrow\infty$. To accommodate the presence of nuisance parameters, decompose
$\thetabf$ as $\thetabf=(\psibf^{T},\phibf^{T})^{T}$, where $\phibf\in\real{p}$, $p\leq m$, is the parameter vector of interest and $\psibf\in\real{m-p}$
is the nuisance parameter vector. The asymptotic covariance matrix of $\phibft$ is represented as $\V_{\phibfs\phibfs}(\thetabf)$, which is the $p\times p$ lower right block of $\V(\thetabf)$.
 Then Cook and Zhang (2015a) defined the envelope for improving $\phibft$ as the smallest reducing subspace of $\V_{\phibfs\phibfs}(\thetabf)$ that contains $\spn(\phibf)$,  $\espc_{\V_{\phibfs\phibfs}(\thetabfs)}(\spn(\phibf))\subseteq\real{p}$.
This definition  links the envelope to a particular
pre-specified method of estimation through the covariance matrix $\V_{\phibfs\phibfs}(\thetabf)$,
while normal-theory maximum likelihood is the only method of estimation allowed by the previous approaches. The goal of an envelope is to improve that pre-specified
estimator, perhaps a maximum likelihood, least squares or robust estimator. Second, the matrix
to be reduced -- here $\V_{\phibfs\phibfs}(\thetabf)$ -- is
dictated by the method of estimation. Third, the matrix to be reduced
can now depend on the parameter being estimated, in addition to perhaps
other parameters.  Cook and Zhang (2015a) sketched application details for generalized linear models, weighted least squares, Cox regression and described an extension to matrix-valued parameters.

\subsection{Computational issues}\label{sec:compissues}

The approaches reviewed in the last section all require estimation of an envelope, now represented generically as $\semu$, the smallest reducing subspace of $\M \in \sreal{r \times r}$ that contains $\uspc \subseteq \real{r}$, where $\M > 0$.  Let $u = \dim(\semu)$, let $\Gammabf \in \real{r \times u}$ be a semi-orthogonal basis matrix for $\semu$,  let $\Mhat$ be a $\sqrt{n}$-consistent estimator of $\M$, and let $\Uhat$ be a positive semi-definite $\sqrt{n}$-consistent estimator of a basis matrix $\U$ for $\uspc$.  With $u$ specified, the most common objective function used for envelope estimation is  
\begin{equation}\label{LG}
L_{u}(\G) = \log |\G^{T}\Mhat\G| +  \log | \G^{T} (\Mhat + \Uhat)^{-1}\G|,
\end{equation}
and the envelope is estimated as $\semuhat = \spn\{ \arg \min L_{u}(\G)\}$, where the minimum is taken over all semi-orthogonal matrices $\G \in \real{r \times u}$. 
Objective function (\ref{LG}) corresponds to maximum likelihood estimation under normality for many envelopes, including those associated with (\ref{mlm}).  Otherwise it provides a $\sqrt{n}$-consistent estimator of the projection onto $\semu$ provided $\Mhat$ and $\Uhat$ are $\sqrt{n}$-consistent (Cook and Zhang, 2015c, who also provided additional background on $L_{u}(\G)$).

 In the case of response envelopes reviewed in Section~\ref{sec:review}, $\Mhat$ is the covariance matrix of the residuals from the ordinary least squares fit of (\ref{mlm}), denoted $\Sbf_{\Y\mid \X}$, and $\Mhat + \Uhat$ is marginal sample covariance matrix of $\Y$, denoted $\Sbf_{\Y}$, and the envelope estimator $\betabfhat = \Pbf_{\espchat} \B$ is the maximum likelihood estimator if the errors are normal.  If the errors are not normal but have finite fourth moments then $\betabfhat$ is $\sqrt{n}$-consistent and asymptotically normal. In the general  context of Cook and Zhang (2015a), also reviewed in Section~\ref{sec:review}, $\Mhat$ is set to a $\sqrt{n}$-consistent estimator of   $\V_{\phibfs\phibfs}(\thetabf)$ and $\Uhat = \phibft \phibft^{T}$.

 For any orthogonal matrix $\Obf \in \real{u \times u}$, $L_{u}(\G) = L_{u}(\G\Obf)$, so  $L_{u}(\G)$ depends only on $\spn(\G)$ and not on a particular basis.  Thus the optimization problem is over a Grassmannian (See Edelman et al. (1998) for background on optimization over Grassmann manifolds.).  Since it takes $u(r-u)$ real numbers to specify $\semu$ uniquely, Grassmann optimization is usually computationally straightforward when $u(r-u)$ is not too large, but it can be very slow when $u(r-u)$ is large.  Also, since $L_{u}(\G)$ is non-convex, the solution  returned may correspond to a local rather than global minimum, particularly when the signal is small relative to the noise.   

It is important that we have a fast and reliable method of determining $\arg \min L_{u}(\G)$ because we may need to repeat that operation hundreds or even thousands of times in an analysis.  An information criterion like AIC or BIC is often used to select a suitable value for $u$, and this requires that we find $\arg \min L_{u}(\G)$ for $u = 0,1,\ldots,r$.  Predictive cross validation might also be used to select $u$, again requiring many optimizations of $L_{u}(\G)$; repeating five fold cross validation with 50 random partitions require in total $250 \times r$ optimizations.  Asymptotic standard errors are available for many normal models, but we may wish to use a few hundred bootstrap samples to determine standard errors when normality is in doubt or when we wish to check the accuracy of the asymptotic approximations.  And may more bootstrap samples may be required if we want accurate inference statements.  In some analyses we may wish to fit a few model variations, again multiplying the computation time.  In cases like those discussed at the end of Section~\ref{sec:review}, $\M = \V_{\phibfs\phibfs}(\thetabf)$, which may depend on unknown parameters, necessitating another level of iteration for the best results (See Cook and Zhang 2015a for further discussion of this point.)  In short, a seemingly small savings in computation time for one optimization of $L_{u}(\G)$ can translate into massive savings over the course of an analysis.  Additionally, the choice of starting value for $\G$ can be crucial since the objective function is non-convex.  Converging to a local minimum can negate the advantages of maximum likelihood estimation, for example.  Trying several different starting values is not really an effective method since it again multiplies the total computation time and in our experience is not likely to result in the global optimum.

Cook, Su and Yang (2014; https://github.com/emeryyi/envlp) developed a fairly comprehensive Matlab toolbox \textsl{envlp} for envelope estimation based on Lippert's  \textsl{sg\_min} program for optimization over Stiefel and Grassmann manifolds  (http://web.mit.edu/$\sim$~ripper/www/software/). This is a very effective toolbox for small to moderate sized analyses, but  otherwise is susceptible to all of the issues mentioned previously. Cook and Zhang (2015c) replaced $L_{u}(\G)$ with a sequential {\em 1D algorithm} that can be computationally much faster than \textsl{sg\_min} and is less dependent on good starting values.  Nevertheless, it is still susceptible to the problems described previously, although less so than methods based on \textsl{sg\_min}.  Additionally, since it does not provide $\arg \min L_{u}(\G)$, it loses the advantages of that accrue with maximum likelihood estimation when normality is a reasonable assumption.  For instance, information criteria like AIC and BIC are no longer available straightforwardly, and likelihood ratio testing is problematic and thus dimension selection must typically be guided by cross validation.

In this paper we propose an iterative non-Grassmann method to compute  $\arg \min L_{u}(\G)$ that is faster and more reliable that existing methods in large analyses and otherwise performs about the same.  It depends crucially on new effective $\sqrt{n}$-consistent starting values that can also be used as standalone estimators.  We restrict our comparisons to the 1D algorithm, since Cook and Zhang (2015c) have demonstrated its superiority over direct optimization methods based on \textsl{sg\_min}.

The new starting values are developed  in Section~\ref{sec:start} and the new algorithm, which relies the new starting values, is described in Section~\ref{sec:algorithm}.
Supporting simulation results are given in Section~\ref{sec:simulations} and contrasts on real data are given in Section~\ref{sec:data}.  Proofs are given in an appendix.

\section{Starting values}\label{sec:start}

In this section we describe how to choose the $u$ columns of the starting value for $\G$ from the eigenvectors of $\Mhat$ or $\Mhat + \Uhat$. To gain intuition about the approach, consider the population representations $\U = \Gammabf \V\Gammabf^{T}$, $\M = \Gammabf \Omegabf \Gammabf^{T} + \Gammabf_{0} \Omegabf_{0} \Gammabf_{0}^{T}$ and 
$(\M + \U)^{-1} = \Gammabf (\Omegabf + \V)^{-1}\Gammabf^{T} + \Gammabf_{0} \Omegabf_{0}^{-1} \Gammabf_{0}^{T}$.  For the starting values selected from the eigenvectors of $\Mhat$  to work well,  the eigenvalues of $\Omegabf$ need to be well distinguished from those of $\Omegabf_{0}$.  If some of the eigenvalues of $\Omegabf$ are close to a subset of the eigenvaues of  $\Omegabf_{0}$ then in samples the corresponding eigenspaces will likely be confused when attempting to minimize $L_{u}(\G)$.  In other words, we may well pick vectors near $\spn(\Gammabf_{0})$ instead of eigenvectors near $\spn(\Gammabf) = \semu$.  In such cases we may obtain a better starting value by choosing from the eigenvectors of $\Mhat+ \Uhat$ rather than the eigenvectors of $\Mhat$.  The same argument applies to choosing the starting values from the eigenvectors of $\Mhat+ \Uhat$: the eigenvalues of $\Omegabf + \V$ need to be well distinguished from those of $\Omegabf_{0}$.  If some of the eigenvalues of $\Omegabf + \V$ are close to a subset of the eigenvalues of  $\Omegabf_{0}$ then in samples the corresponding eigenspaces will again likely be confused.  In such cases we may obtain better starting values by starting with the eigenvectors of $\Mhat$ rather than the eigenvectors of $\Mhat+ \Uhat$.    The general conclusion from this discussion is that for effective starting values we will  need to consider both $\Mhat$ and $\Mhat + \Uhat$. Scaling will also be an issue, as discussed later in this section, leading to four potential starting values.  The actual starting value used is the one that minimizes $L_{u}(\G)$.

We make use of the following result.

\begin{proposition}\label{prop:min1}
Let $(\G, \G_{0}) \in \real{r \times r}$ be an orthogonal matrix with $\G \in \real{r \times u}$ and let $\M \in \sreal{r \times r}$ be a positive definite matrix.  Then $\log | \G^{T}\M\G | + \log |\G_{0}^{T}\M\G_{0}|$ and $\log | \G^{T}\M\G | + \log |\G^{T}\M^{-1}\G|$ are both minimized globally when the columns of $\G$ span any $u$ dimensional reducing subspace of $\M$.
\end{proposition}

 In the next  section we describe how to select starting values from the eigenvectors of $\Mhat$.
\subsection{Choosing the starting value from the eigenvectors of $\Mhat$}\label{sec:M}

Define $J_{1}(\G) =  \log|\G^T\Mhat\G| + \log |\G_{0}^T\Mhat\G_{0}| $, $J_{2}(\G) = \log | \I_{r-u} +\G_{0}^{T}\Uhat_{\M}\G_{0}|$ and $J(\G) = J_{1}(\G) + J_{2}(\G)$, where $\Uhat_{\M} = \Mhat^{-1/2}\Uhat\Mhat^{-1/2}$ is a standardized version of $\Uhat$.  Assume for convenience that the eigenvalues of $\Mhat$ are unique, which will typically hold with probability $1$, and let $\vspc_{u}$ be the collection of all subsets of $u$ eigenvectors of $\Mhat$. Then
\begin{proposition}\label{prop:min2}
 $\arg \min_{\G \in \vspc_{u} }L_{u}(\G) = \arg \min_{\G \in \vspc_{u}}J(\G)$. 
\end{proposition}
Consequently, instead of $L_{u}(\G)$ we can work with the more amenable objective function $J(\G)=J_{1}(\G) + J_{2}(\G)$ when restricting starting values to the eigenvectors of $\Mhat$.   It follows from Proposition~\ref{prop:min1} that $J_{1}(\G)$ is minimized  when the columns of $\G$ are any $u$ eigenvectors of $\Mhat$.    Restricting $\G \in \vspc_{u}$, we next need to find $\arg \min_{\G \in \vspc_{u}} J_{2}(\G)$.   This does not have a closed-form solution and  evaluating at all $r$-choose-$u$ elements of $\vspc_{u}$  will be effectively impossible when $r$ is large.  For these reasons we replace the $\log$-determinant in $J_{2}(\G)$ with the trace and minimize $ \tr(\I_{r-u} + \G_{0}^{T}\Uhat_{\M}\G_{0}) $, which is equivalent to maximizing 
\[
K_{\M}(\G) := \tr(\G^{T}\Uhat_{\M}\G)  = \sum_{i=1}^{u} \g_{i}^{T}\Uhat_{\M}\g_{i},
\]
where $\g_{i}$ is the $i$-th selected eigenvector of $\Mhat$ (the $i$-th column of $\G$).  Computation is now easy, since we just select the $u$ eigenvectors of $\Mhat$ that maximize $\g_{i}^{T}\Uhat_{\M}\g_{i} $.

Applying this in response envelopes, let $\Sbf_{\X}$ denote the marginal sample covariance matrix of the predictors.  Then $\Mhat = \Sbf_{\Y\mid \X}$, $\Uhat = \B \Sbf_{\X}\B^{T}$, $\Uhat_{\M} = \Sbf_{\Y\mid \X}^{-1/2}\B \Sbf_{\X}\B^{T}\Sbf_{\Y\mid \X}^{-1/2}$, and   $\Sbf_{\Y\mid \X}^{-1/2}\B \Sbf_{\X}^{1/2}$ is a standardized version of the ordinary least squares estimator $\B$ of $\betabf$.

%

\subsection{Choosing the starting value from the eigenvectors of $\Mhat + \Uhat$}\label{sec:M+U}

Define $J_{1}^{*}(\G) =  \log|\G^T(\Mhat+\Uhat)\G| + \log |\G^T(\Mhat+\Uhat)^{-1}\G|$, $J_{2}^{*}(\G) = \log | \I_{u} -\G^{T}\Uhat_{\M+\U}\G|$ and $J^{*}(\G) = J_{1}^{*}(\G) + J_{2}^{*}(\G)$, where $\Uhat_{\M+\U} = (\Mhat+\Uhat)^{-1/2}\Uhat(\Mhat+\Uhat)^{-1/2}$ is another standardized version of $\Uhat$.  Let $\vspc_{u}^{*}$ be the collection of all subsets of $u$ eigenvectors of $\Mhat + \Uhat$. Then
\begin{proposition}\label{prop:min3}
 $\arg \min_{\G \in \vspc_{u} ^{*}}L_{u}(\G) = \arg \min_{\G \in \vspc_{u}^{*}}J^{*}(\G)$. 
\end{proposition}

Consequently, instead of $L_{u}(\G)$ we can again work with a more amenable objective function, this time $J^{*}(\G)=J_{1}^{*}(\G) + J_{2}^{*}(\G)$.   It follows from Proposition~\ref{prop:min1} that $J_{1}^{*}(\G)$ is minimized  when the columns of $\G$ are any $u$ eigenvectors of $\Mhat+\Uhat$.   Restricting $\G \in \vspc_{u}^{*}$, we next need to find $\arg \min_{\G \in \vspc_{u}} J_{2}^{*}(\G)$.   Again, this does not have a closed-form solution and  evaluating at all $r$-choose-$u$ elements of $\vspc_{u}^{*}$  will be effectively impossible when $r$ is large.  For these reasons we again replace the $\log$-determinant with the trace and minimize $ \tr(\I_{u} - \G^{T}\Uhat_{\M+\U}\G) $, which is equivalent to maximizing 
\[
K_{\M+\U}(\G) := \tr(\G^{T}\Uhat_{\M+\U}\G)  = \sum_{i=1}^{u} \g_{i}^{T}\Uhat_{\M+\U}\g_{i},
\]
where $\g_{i}$ is the $i$-th selected eigenvector of $\Mhat+\Uhat$ (the $i$-th column of $\G$).  Computation is again easy, since we just select the $u$ eigenvectors of $\Mhat+\Uhat$ that maximize $\g_{i}^{T}\Uhat_{\M+\U}\g_{i} $.  This is exactly the same as the previous case, except the standardization of $\Uhat$ is with  $(\Mhat + \Uhat)^{-1/2}$ instead of $\Mhat^{-1/2}$.
  
  Applying this in response envelopes, $\Mhat = \Sbf_{\Y\mid \X}$, $\Uhat = \B \Sbf_{\X}\B^{T}$, $\Mhat + \Uhat = \Sbf_{\Y}$, $\Uhat_{\M+\U} = \Sbf_{\Y}^{-1/2}\B \Sbf_{\X}\B^{T}\Sbf_{\Y}^{-1/2}$ and $ \Sbf_{\Y}^{-1/2}\B \Sbf_{\X}^{1/2}$ is just a standardized matrix of ordinary least squares regression coefficients as before.

  \subsection{Scaling and  consistency}
  
   The standardized forms $\Uhat_{\M}$ and $\Uhat_{\M+\U}$ are important when the scales involved in $\Mhat$ and $\Mhat + \Uhat$ are very different.  This can perhaps be appreciated readily in the context of response envelopes, where $\Mhat = \Sbf_{\Y \mid \X}$ and $\Mhat + \Uhat = \Sbf_{\Y}$.  In this case the standardization will be important and effective if the scales of the elements of $\Y$ are very different. However, the standardization will be effectively unnecessary when the scales are similar.  In the case of response envelopes this means that the scales of the elements of $\Y$ are the same or similar.  
   
   Depending on the scales involved, standardization can also be counterproductive when the sample size is not large enough to give sufficiently accurate estimates of $\M$ and $\U$.  In such cases, we abandon the standardization and use either 
$K_{\M}^{*}(\G) = \sum_{i=1}^{u} \g_{i}^{T}\Uhat\g_{i}$
or 
$
K_{\M+\U}^{*}(\G) = \sum_{i=1}^{u} \g_{i}^{T}\Uhat\g_{i}
$
as the objective function.  The only difference between these is that $K_{\M}^{*}(\G)$ confines $\G$ to the eigenvectors of $\Mhat$, while $K_{\M+\U}^{*}(\G)$ confines $\G$ to the eigenvectors of $\Mhat + \Uhat$.  We now have four possible starting values from which to choose, corresponding to the arguments that minimize $K_{\M}(\G)$, $K_{\M}^{*}(\G)$, $K_{\M+\U}(\G)$, and $K_{\M+\U}^{*}(\G)$. The value $\G_{\start}$ chosen to start the algorithm described in Section~\ref{sec:algorithm} is the one that minimizes $L_{u}(\G)$.

We conclude this section with the following consistency result:
\begin{proposition}\label{prop:consistency}
Let $\Pbf_{\start}$ denote the projection onto $\spn(\G_{\start})$.  Then with known $u$, $\Pbf_{\start}$ is a $\sqrt{n}$-consistent of the projection onto $\semu$.
\end{proposition}

\section{New iterative algorithm}\label{sec:algorithm}

In this section we describe a re-parameterized version of $L_{u}(\G)$ that does not require optimization over a Grassmannian.  The new parameterization requires first selecting $u$ rows of $\G \in \real{r \times u}$ and then constraining the matrix $\G_{1} \in \real{u \times u}$ formed with these rows to be non-singular.    
Without loss of generality, assume that $\G_{1}$ is constructed from the first $u$ rows of $\G$ which we can then partition as 
 \[
 \G = \left( \begin{array} {c} \G_{1} \\ \G_{2} \end{array} \right) = \left( \begin{array} {c}\I_{u} \\ \A \end{array} \right)\G_{1} = \C_{\A}\G_{1},
  \]
 where $\A = \G_{2}\G_{1}^{-1} \in \real{(r-u)\times u }$ is an unconstrained matrix and $\C_{\A} = (\I_{u}, \A^{T})^{T}$.  Since $\G^{T}\G = \I_{u}$ and $\G_{1}$ is non-singular, $\G_{1}\G_{1} ^{T}=  (\C_{\A}^{T}\C_{\A})^{-1}$. 
 Using these relationships, $L_{u}(\G)$ can be re-parameterized as a function of only $\A$: 
  \[
 L_{u}(\A) = -2 \log |\C_{\A}^{T}\C_{\A}| + \log\left|  \C_{\A}^{T}\Mhat \C_{\A} \right| + \log\left|  \C_{\A}^{T}(\Mhat + \Uhat)^{-1}\C_{\A} \right|.
 \]
 With this objective function minimization over $\A$ is unconstrained.  The number of real parameters $u(r-u)$ comprising $\A$ is the same as the number of reals needed to specify a single element in the Grassmannian $\gspc(u,r)$.
 
 If $u(r-u)$ is not too large, $L(\A)$ might be minimized directly by using standard optimization software and the starting values described in Section~\ref{sec:start}.  In other cases minimization can be carried out by minimizing iteratively over the rows of $\A$.  Suppose that we wish to minimize over the last row $\abf^{T}$ of $\A$.  Partition
 \[
 \A = \left(\!\!\begin{array}{c} \A_{1} \\ \abf^{T} \end{array} \!\!\right), \;\C_{\A} = \left(\!\!\begin{array}{c} \C_{\A_{1}} \\ \abf^{T} \end{array} \!\!\right),\; \Mhat = \left(\!\begin{array}{cc} \Mhat_{11} & \Mhat_{12}\\ \Mhat_{21} & \widehat{M}_{22} \end{array}\! \!\right), \; (\Mhat + \Uhat)^{-1}= \left(\!\begin{array}{cc} \Vhat_{11} & \Vhat_{12}\\ \Vhat_{21} & \widehat{V}_{22} \end{array} \!\!\right).
 \]
 Then after a little algebra, the objective function for minimizing over $\abf^{T}$ with $\A_{1}$ held fixed can be written up to terms that do not depend on $\abf$ as
\begin{eqnarray*}
L_{u}(\abf \mid \A_{1})  &=& -2\log \left\{1+\abf^{T}(\C_{\A_{1}}^{T}\C_{\A_{1}})^{-1}\abf\right\} \\
& & +\log\left\{1+\widehat{M}_{22}(\abf+\widehat{M}_{22}^{-1}\C_{\A_{1}}^{T}\Mhat_{12})^{T}\W_{1}^{-1}(\abf+\widehat{M}_{22}^{-1}\C_{\A_{1}}^{T}\Mhat_{12}) \right\}\\
& &+\log\left\{1+\widehat{V}_{22}(\abf+\widehat{V}_{22}^{-1}\C_{\A_{1}}^{T}\Vhat_{12})^{T}\W_{2}^{-1}(\abf+\widehat{V}_{22}^{-1}\C_{\A_{1}}^{T}\Vhat_{12}) \right\},
\end{eqnarray*}
where
\begin{eqnarray*}
\W_{1}  &=& \C_{\A_{1}}^{T}\left(\Mhat_{11} - \widehat{M}_{22}^{-1} \Mhat_{12} \Mhat_{21}\right)\C_{\A_{1}} \\
\W_{2}  &=& \C_{\A_{1}}^{T}\left(\Vhat_{11} - \widehat{V}_{22}^{-1} \Vhat_{12} \Vhat_{21}\right)\C_{\A_{1}}.
\end{eqnarray*}
The objective function $L(\abf \mid \A_{1})$ can now be minimized using any suitable off-the-shelf algorithm.  Iteration then cycles over rows of $\A$ until a convergence criterion is met.

This algorithm requires the starting value $\G_{\start}$ described in Section~\ref{sec:start}.  Prior to application of the algorithm we must identify $u$ rows of $\G_{\start}$ and then constrain the matrix $\G_{\start,u}$ formed from those $u$ rows to be non-singular.  
This implies that the  matrix  formed from the corresponding rows of a basis matrix for $\espc_{\M}(\uspc)$ should also be non-singular. 
This can be achieved reliably by first applying Gaussian elimination with partial pivoting to $\G_{\start}$.    The $u$ rows of $\G_{\start}$ identified during this process then form $\G_{\start,u}$.

\begin{proposition}\label{prop:ge}
Assume that the eigenvalues of $\M$ and $\M+\U$ are distinct.  
Then the $u \times u$ submatrix of $\G_{\start}$ that consists of the $u$ rows selected by Gaussian elimination converges to a non-singular matrix with rate $\sqrt{n}$.
\end{proposition} 
This proposition shows that asymptotically Gaussian elimination produces a non-singular submatrix.  The condition that the eigenvalues of $\M$ and $\M+\U$ be distinct is mainly for clarity of exposition and is not necessary.  The  proof given in the appendix demonstrates a more complete result.   Let $\Gammabf_{\start}$ denote the population version of $\G_{\start}$, and let $\Gammabf_{\start, u} \in \real{u \times u}$ consist of the $u$ rows of $\Gammabf_{\start}$ formed by applying Gaussian elimination to $\Gammabf_{\start}$.  Then $\Gammabf_{\start}$ is a basis matrix for $\semu$ and $\G_{\start, u}$ converges to $\Gammabf_{\start, u}$ at rate $\sqrt{n}$. 
%

 The new algorithm estimates a basis $\Gammabf$ row by row, while the 1D algorithm optimizes column by column.  When $u$ is small, the 1D algorithm tends to be a bit more efficient as it optimizes one column at a time and it needs only one pass through those columns.    When $u$ is larger, the new algorithm dominates, and sometimes substantially.  In each estimation, the 1D algorithm uses conjugate gradient with Polak-Ribiere updates while our algorithm uses Newton updates.   

\section{Simulations}\label{sec:simulations}

\subsection{Starting values}\label{sec:startsim}

 The first series of simulations was designed to illustrate why it is important to consider the eigenvalues of both $\Mhat$ and $\Mhat+\Uhat$.   All simulations are for response envelopes reviewed in Section~\ref{sec:review},  model (\ref{envmodel}).   The results displayed in the tables of this section are the average over 50 replications in each simulation scenario.  The angle $\angle(\spn(\A_{1}), \spn(\A_{2}))$ between the subspaces spanned by columns of the semi-orthogonal basis matrices $\A_{1} \in \real{r \times u}$ and $\A_{2} \in \real{r \times u}$  was computed in degrees as the arc cosine of the smallest absolute singular value of $\A^{T}\B$, and $\betabfhat_{\start} = \Pbf_{\start}\B$, where $\Pbf_{\start}$ is as defined in Proposition~\ref{prop:consistency}.  The starting value is still denoted as $\G_{\start}$ but its definition depends on the simulation. $\Gammabfhat = \arg \min L(\G)$ was obtained from the new algorithm described in Section~\ref{sec:algorithm} using the simulation-specific starting value $\G_{\start}$, and $\semuhat = \spn(\Gammabfhat)$.   

\paragraph{Scenario I.}

 This simulation was designed to illustrate a regression in which the eigenvalues of $\Sigmabf$ are close and the signal is strong. We generated the data with $p=r=100$, $n=500$ and $u=20$, taking $\Omegabf$ and $\Omegabf_{0}$ to be diagonal matrices with diagonal elements generated as independent uniform $(49, 51)$ variates.  Elements in $\etabf$ were independent 
 uniform $(0, 10)$ variates, $\X$ followed a multivariate normal distribution with mean $0$ and covariance matrix $400\I_{p}$, and the elements of $(\Gammabf, \Gammabf_{0}) \in \real{r \times r}$ were obtained 
 by standardizing a matrix of independent uniform $(0, 1)$ variates.    In this scenario, the eigenvalues of $\Sigmabf$ are close to each other, but we have a strong signal arising from the distribution of $\X$.  Starting values based on the eigenvectors of $\Mhat = \Sbf_{\Y|\X}$ were expected to perform poorly, while starting values based on $\Mhat+\Uhat = \Sbf_{\Y}$ were expected to perform well, as conjectured at the start of Section~\ref{sec:start} and confirmed by the results in Table \ref{tab:scenarioI}.

\begin{table}[ht!]
\centering
\renewcommand{\arraystretch}{1.25}
\begin{tabular}{ |c | c c | c c|}
\hline
 & \multicolumn{2}{c |}{Standardized $\Uhat$} &  \multicolumn{2}{c |}{Unstandardized $\Uhat$}\\
Summary statistic    &  $\Sbf_{\Y} = \Mhat + \Uhat$ & $\Sbf_{\Y|\X} = \Mhat$ & $\Sbf_{\Y} = \Mhat + \Uhat$ & $\Sbf_{\Y|\X} = \Mhat$\\
   \hline\hline
$\angle (\spn(\G_{\start}), \semu)$   & $0.58$ & $89.05$ & $0.58$ & $88.98$\\   \hline
$\angle(\semuhat, \semu)$ & $0.58$ & $88.58$ & $0.58$ & $88.74$\\   \hline
$L_{u}(\G_{\start})$   & $-182.10$ & $-13.18$ & $-182.10$ & $-9.94$\\   \hline
$L_{u}(\Gammabfhat)$   & $-182.10$ & $-21.95$ & $-182.10$ & $-20.01$\\   \hline
 $\|\betabfhat_{\start}-\betabf\|_{2}$    & $0.27$ & $149.58$ & $0.27$ & $136.51$\\   \hline
 $\|\betabfhat-\betabf\|_{2}$  & $0.27$ & $113.02$ & $0.27$ & $101.67$\\   \hline
     \end{tabular}
\caption{Results for Scenario I.  The starting value $\G_{\start}$ was constructed from the eigenvectors of the matrices indicated by the headings for columns 2-5. }\label{tab:scenarioI}
\end{table}
The overarching conclusion from Table \ref{tab:scenarioI} is that the starting values from $\Sbf_{\Y}$ did very well, whether  $\Uhat$ was standardized or not, while the starting values from $\Sbf_{\Y|\X}$ were effectively equivalent to choosing a 20-dimensional subspace at random.  Additionally, iteration from the starting value produced essentially no change in the angle, the value of the objective function or the envelope estimator of $\betabf$.

\paragraph{Scenario II.}

We generated data with $p=r=100$, $n=500$ and $u=5$, taking  $\Omegabf=\I_{u}$ and $\Omegabf_{0}=100\I_{r-u}$.  Elements in $\etabf$ were independent uniform $(0, 10)$ variates, $\X$ followed multivariate normal distribution with mean $0$ and covariance matrix $25\I_{p}$, and $(\Gammabf, \Gammabf_{0})$ was obtained by standardizing an $r\times r$ matrix of independent uniform $(0, 1)$ variates.  Since the eigenvalues in $\Omegabf$ and $\Omegabf_{0}$ are very different and the signal is modest, the results in Table~\ref{tab:scenarioII} show as expected from the argument given in Section 2 that the starting values based on $\Mhat = \Sbf_{\Y|\X}$ did much better than those based on $\Sbf_{\Y}$.  As in Scenario I, the starting value did very well.  Iteration improved the starting value a small amount and scaling had no notable affect.
\begin{table}[ht!]
\centering
\renewcommand{\arraystretch}{1.25}
\begin{tabular}{ |c | c c | c c|}
\hline
 & \multicolumn{2}{c |}{Standardized $\Uhat$} &  \multicolumn{2}{c |}{Unstandardized $\Uhat$}\\
Summary statistic    &  $\Sbf_{\Y} = \Mhat + \Uhat$ & $\Sbf_{\Y|\X} = \Mhat$ & $\Sbf_{\Y} = \Mhat + \Uhat$ & $\Sbf_{\Y|\X} = \Mhat$\\
   \hline\hline
$\angle (\spn(\G_{\start}), \semu)$   & $45.88$ & $3.87$ & $45.88$ & $3.87$\\   \hline
$\angle(\semuhat, \semu)$  & $36.55$ & $3.78$ & $36.55$ & $3.78$\\   \hline
$L_{u}(\G_{\start})$    & $-16.19$ & $-30.88$ & $-16.19$ & $-30.88$\\   \hline
$L_{u}(\Gammabfhat)$   & $-20.74$ & $-30.95$ & $-20.74$ & $-30.95$\\   \hline
 $\|\betabfhat_{\start}-\betabf\|_{2}$   & $1.93$ & $0.66$ & $1.93$ & $0.66$\\   \hline
 $\|\betabfhat-\betabf\|_{2}$  & $1.64$ & $0.57$ & $1.64$ & $0.57$\\   \hline
     \end{tabular}
\caption{Results for scenario II. The starting value $\G_{\start}$ was constructed from the eigenvectors of the matrices indicated by the headings for columns 2-5.}\label{tab:scenarioII}
\end{table}

\paragraph{Scenario III.}

The intent of this simulation is to demonstrate the importance of scaling $\Uhat$.  We generated  data with $p=r=30$, $n=200$ and $u=5$, taking $\Omegabf$ to be a diagonal matrix with diagonal elements $1.5^{1}, \cdots,1.5^{u}$ and $\Omegabf_{0}$ to be a diagonal matrix with diagonal elements $1.5^{u+1}$, $\cdots$, $1.5^{r}$.  Elements in $\etabf$ were generated as independent uniform $(0, 10)$ variates, $\X$ follows the multivariate normal distribution with mean $0$ and covariance matrix $100\I_{p}$, and $(\Gammabf, \Gammabf_{0})=\I_{r}$.     We see from the results of Table~\ref{tab:scenarioIII}  that standardization performs well and that now iteration improves the starting value considerably.  Here and in the other results of this section,  the smallest value of $L_{u}(\G_{\start})$ produced best results.

\begin{table}[ht!]
\centering
\renewcommand{\arraystretch}{1.25}
\begin{tabular}{ |c | c c | c c|}
\hline
 & \multicolumn{2}{c |}{Standardized $\Uhat$} &  \multicolumn{2}{c |}{Unstandardized $\Uhat$}\\
Summary statistic    &  $\Sbf_{\Y} = \Mhat + \Uhat$ & $\Sbf_{\Y|\X} = \Mhat$ & $\Sbf_{\Y} = \Mhat + \Uhat$ & $\Sbf_{\Y|\X} = \Mhat$\\
   \hline\hline
$\angle (\spn(\G_{\start}), \semu)$  & $48.63$ & $16.72$ & $89.35$ & $33.31$\\   \hline
$\angle(\semuhat, \semu)$ & $17.92$ & $1.54$ & $89.34$ & $22.77$\\   \hline
$L_{u}(\G_{\start})$    & $-13.43$ & $-35.75$ & $-12.69$ & $-34.09$\\   \hline
$L_{u}(\Gammabfhat)$   & $-32.48$ & $-46.93$ & $-23.26$ & $-44.84$\\   \hline
 $\|\betabfhat_{\start}-\betabf\|_{2}$   & $11.82$ & $8.56$ & $20.32$ & $11.13$\\   \hline
 $\|\betabfhat-\betabf\|_{2}$  & $4.37$ & $0.72$ & $20.17$ & $5.39$\\   \hline
     \end{tabular}
\caption{Results for scenario III. The starting value $\G_{\start}$ was constructed from the eigenvectors of the matrices indicated by the headings for columns 2-5.}\label{tab:scenarioIII}
\end{table}

\paragraph{Scenario IV.}  For this simulation we kept the same settings as Scenario III, except that diagonal elements of $\Omegabf$ and $\Omegabf_{0}$ were $1.05^{1}$, $\cdots$, $1.05^{u}$ and $1.05^{u+1}$, $\cdots$, $1.05^{r}$, and $(\Gammabf, \Gammabf_{0})$ was generated by standardizing a matrix of uniform $(0, 1)$ random variables. In this setup heteroscedasticity across the elements is reduced substantially from that in scenario III.   As indicated in Table \ref{tab:scenarioIV}, the standardization no longer provides much improvement. Also, since the eigenvalues of $\Omega$ and $\Omega_0$ are similar,  $S_{\Y|\X}$ again does not work well. 
\begin{table}[ht!]
\centering
\renewcommand{\arraystretch}{1.25}
\begin{tabular}{ |c | c c | c c|}
\hline
 & \multicolumn{2}{c |}{Standardized $\Uhat$} &  \multicolumn{2}{c |}{Unstandardized $\Uhat$}\\
Summary statistic    &  $\Sbf_{\Y} = \Mhat + \Uhat$ & $\Sbf_{\Y|\X} = \Mhat$ & $\Sbf_{\Y} = \Mhat + \Uhat$ & $\Sbf_{\Y|\X} = \Mhat$\\
   \hline\hline
$\angle (\spn(\G_{\start}), \semu)$  & $0.30$ & $79.57$ & $0.30$ & $80.66$\\   \hline
$\angle(\semuhat, \semu)$ & $0.30$ & $73.40$ & $0.30$ & $75.58$\\   \hline
$L_{u}(\G_{\start})$    & $-53.54$ & $-7.92$ & $-53.54$ & $-7.27$\\   \hline
$L_{u}(\Gammabfhat)$   & $-53.54$ & $-13.40$ & $-53.54$ & $-12.56$\\   \hline
 $\|\betabfhat_{\start}-\betabf\|_{2}$   & $0.08$ & $33.36$ & $0.08$ & $31.59$\\   \hline
 $\|\betabfhat-\betabf\|_{2}$ & $0.08$ & $25.04$ & $0.08$ & $22.61$\\   \hline
     \end{tabular}
\caption{Results for scenario IV. The starting value $\G_{\start}$ was constructed from the eigenvectors of the matrices indicated by the headings for columns 2-5.}\label{tab:scenarioIV}
\end{table}

\subsection{Comparisons with the 1D algorithm}

In this section we give three different simulation scenarios based on response envelopes for comparing the new non-Grassmann algorithm with the 1D algorithm.  In all scenarios $p=100$, $\alphabf = 0$,  orthogonal bases $(\Gammabf, \Gammabf_{0})$ were obtained by normalizing an $r\times r$ matrix of independent uniform $(0, 1)$ variates, the  elements in $\etabf\in\real{u\times p}$ were generated as  independent uniform $(0, 10)$ variates, and $\betabf=\Gammabf\etabf$.   The predictors $\X$ were generated as independent normal random vectors with mean $0$ and variance $400\I_{r}$. We varied $u$ from $1$ to $90$ and recorded and computing times and the angles between the true and estimated subspaces.  

The 1D algorithm was implemented in R for all  simulations reported in this and the next section. Using efficient programming tools in R, it is now much faster than its Matlab version, which produced the results in Cook and Zhang (2015c).
To insure a fair comparison, we used the default convergence criterion in R for optimizations within both the 1D algorithm and the new algorithm.  The angle between subspaces was computed as described previously. 
In all case the results tabled are the averages over 50 replications. We use $\Gammabfhat_{\oneD}$ to denote the basis generated by the the 1D algorithm.

\paragraph{Scenario V.}  In this scenario we set $r=100$ and $n=250$.  To reflect multivariate regressions with large immaterial variation, so envelopes give large gains, we generated the error covariance matrix as $\Sigmabf=\Gammabf\Omegabf\Gammabf^{T}+\Gammabf_{0}\Omegabf_{0}\Gammabf_{0}^{T}$, where $\Omegabf=\A\A^{T}$, $\Omegabf_{0}=\B\B^{T}$, the elements in $\A$ were generated as independent standard normal variates and elements in $\B$ were generated as independent normal $(0, 5^{2})$ variates.  The results are shown in Table~\ref{tab:scenarioV}.

The 1D algorithm tends to perform a bit better on accuracy (Table \ref{tab:scenarioV}) for small values of $u$, while performing poorly for large values of $u$.  The same phenomenon occurs in terms of time:  the 1D algorithm tends to be a bit faster for small values of $u$, but otherwise can take much longer than the new non-Grassmann algorithm.  The relatively small times for the new algorithm at $u=5,10,20,60$ occurred because in those cases the starting value was quite good and little iteration was required.  The same qualitative differences hold when considering the norm between the estimated coefficient matrix and the true value from the simulation.  Note also that the angle for the starting value by itself was often smaller than that for the 1D algorithm.

\begin{table}[ht!]
\centering
\renewcommand{\arraystretch}{1.25}
\begin{tabular}{ |c | c c c |}
\hline
(A) Angle &  $\semuhat$ &$\spn(\G_{\start})$ & $\spn(\Gammabfhat_{\oneD})$ \\
   \hline\hline
 $  u = 1 $   & $0.92$ & $1.68$ & $0.64$ \\   \hline
 $  u = 5 $   & $3.56$ & $3.60$ & $1.81$ \\   \hline
 $  u = 10 $   & $4.67$ & $4.73$ & $4.60$ \\   \hline
 $  u = 20 $   & $5.83$ & $5.84$ & $42.77$ \\   \hline
 $  u = 30 $   & $4.84$ & $6.07$ & $12.37$ \\   \hline
 $  u = 40 $   & $5.59$ & $7.39$ & $6.24$  \\   \hline
 $  u = 50 $   & $6.81$ & $7.62$ & $39.57$ \\   \hline
 $  u = 60 $   & $8.48$ & $8.49$ & $70.37$ \\   \hline
 $  u = 80 $   & $7.61$ & $10.01$ & $25.51$ \\   \hline
 $  u = 90 $   & $7.15$ & $12.04$ & $21.02$ \\   \hline\hline
 (B) Time &  $\semuhat$ &$\spn(\G_{\start})$ & $\spn(\Gammabfhat_{\oneD})$ \\ \hline
 $  u = 1 $   & $2.30$ & $0.03$  & $0.23$\\   \hline
 $  u = 5 $   & $0.19$ & $0.03$  & $1.45$\\   \hline
 $  u = 10 $   & $0.37$ & $0.03$  & $2.71$\\   \hline
 $  u = 20 $   & $0.34$ & $0.03$  & $5.16$\\   \hline
 $  u = 30 $   & $7.49$ & $0.04$ & $6.23$\\   \hline
 $  u = 40 $   & $7.58$ & $0.04$  & $7.30$\\   \hline
 $  u = 50 $   & $5.53$ & $0.05$  & $9.18$\\   \hline
 $  u = 60 $   & $0.97$ & $0.05$  & $10.59$\\   \hline
 $  u = 80 $   & $2.21$ & $0.07$  & $11.07$\\   \hline
 $  u = 90 $   & $1.55$ & $0.08$  & $10.40$\\   \hline
     \end{tabular}
\caption{Scenario V:  (A) Angle between $\semu$ and the indicated subspace. (B)  Computing time in seconds for the indicated subspace.  $\semuhat$, $\spn(\G_{\start})$ and $\spn(\Gammabfhat_{\oneD})$ denote the estimated subspaces by the new non Grassmann algorithm, the starting values described in Section~\ref{sec:start} and the 1D algorithm.}
\label{tab:scenarioV}
\end{table}

\paragraph{Scenario VI.} We again set $r=100$ and $n=250$.  To reflect multivariate regressions with small immaterial variation, so envelopes give worthwhile but relatively modest gains, we generated the error covariance matrix as $\Sigmabf=\Gammabf\Omegabf\Gammabf^{T}+\Gammabf_{0}\Omegabf_{0}\Gammabf_{0}^{T}$, where $\Omegabf=\A\A^{T}$, $\Omegabf_{0}=\B\B^{T}$, the elements in $\A$ were generated as independent normal $(0, 5^{2})$ variates variates and elements in $\B$ were generated as independent standard normal variates.  The results  shown in Table~\ref{tab:scenarioVI} broadly parallel those in Table~\ref{tab:scenarioV} for Scenario V, but now the performance of the new algorithm is stronger, both in terms of accuracy and time.

\begin{table}[ht!]
\centering
\renewcommand{\arraystretch}{1.25}
\begin{tabular}{ |c | c c c|}
\hline
   (A) Angle &  $\semuhat$ & $\spn(\G_{\start})$ & $\spn(\Gammabfhat_{\oneD})$ \\
   \hline\hline
 $  u = 1 $   & $0.32$ & $0.32$ & $0.33$ \\   \hline
 $  u = 5 $   & $0.75$ & $0.75$ & $0.70$ \\   \hline
 $  u = 10 $   & $0.89$ & $0.89$ & $2.94$ \\   \hline
 $  u = 20 $   & $1.13$ & $1.14$ & $21.00$ \\   \hline
 $  u = 30 $   & $1.24$ & $1.24$ & $10.73$ \\   \hline
 $  u = 40 $   & $1.36$  & $1.36$ & $12.68$ \\   \hline
 $  u = 50 $   & $1.40$ & $1.40$ & $16.97$ \\   \hline
 $  u = 60 $   & $1.45$  & $1.45$ & $31.20$ \\   \hline
 $  u = 80 $   & $1.64$ & $1.64$ & $6.67$ \\   \hline
 $  u = 90 $   & $1.14$ & $1.14$ & $4.10$ \\   \hline\hline
 (B) Time &  $\semuhat$ & $\spn(\G_{\start})$ & $\spn(\Gammabfhat_{\oneD})$ \\ \hline 
  $  u = 1 $   & $0.08$ & $0.04$  & $0.30$\\   \hline
 $  u = 5 $   & $0.10$ & $0.03$  & $1.13$\\   \hline
 $  u = 10 $   & $0.18$ & $0.03$  & $2.52$\\   \hline
 $  u = 20 $   & $0.29$ & $0.04$  & $3.82$\\   \hline
 $  u = 30 $   & $0.42$ & $0.04$  & $6.42$\\   \hline
 $  u = 40 $   & $0.71$ & $0.04$  & $7.71$\\   \hline
 $  u = 50 $   & $0.38$ & $0.05$ & $9.74$\\   \hline
 $  u = 60 $   & $0.31$ & $0.06$  & $10.62$\\   \hline
 $  u = 80 $   & $0.61$ & $0.07$  & $11.69$\\   \hline
 $  u = 90 $   & $0.21$ & $0.09$ & $11.00$\\   \hline
     \end{tabular}
     
\caption{Scenario VI: (A) Angle between $\semu$ and the indicated subspace. (B)  Computing time in seconds for the indicated subspace.  $\semuhat$, $\spn(\G_{\start})$ and $\spn(\Gammabfhat_{\oneD})$ denote the estimated subspaces by the new non Grassmann algorithm, the starting values described in Section~\ref{sec:start} and the 1D algorithm.}
\label{tab:scenarioVI}
\end{table}

\paragraph{Scenario VII.}

This scenario was designed to emphasize the time differences between the 1D algorithm and the non Grassmann algorithm.   We set $n=500$ and varied $r$ from $150$ to $350$.  The error covariance matrix was constructed as $\Sigmabf=\Gammabf\Omegabf\Gammabf^{T}+\Gammabf_{0}\Omegabf_{0}\Gammabf_{0}^{T}$, where $\Omegabf=\I$, $\Omegabf_{0}=25\I$.   The  results in Table~\ref{tab:scenarioVII}A indicate that estimative performance of the two algorithms is essentially the same.  However, as shown in Table~\ref{tab:scenarioVII}B the 1D algorithm can take considerably longer than the non Grassmann algorithm.  To emphasize the differences, the 1D algorithm with $r=350$ would take about $2.5$ hours to estimate the envelope for each $u$ between $1$ and $90$, while the non Grassmann algorithm would take only about $0.15$ hours.  In practice we would normally need to estimate the envelope for each $u$ between $1$ and $350$, leading to much longer computing times.

\begin{table}[ht!]
\centering
\renewcommand{\arraystretch}{1.25}
\begin{tabular}{ | c | c c | c c | cc | }
\hline
 (A) & \multicolumn{2}{c |}{$r=150$} &  \multicolumn{2}{c |}{$r=250$} &  \multicolumn{2}{c |}{$r=350$} \\
 Angle & $\semuhat $& $\spn(\Gammabfhat_{\oneD})$ & $\semuhat $& $\spn(\Gammabfhat_{\oneD})$ & $\semuhat $& $\spn(\Gammabfhat_{\oneD})$\\
   \hline\hline
 $  u = 1 $   & $0.29$  & $ 0.29 $ & $0.34$ & $ 0.34 $ & $0.42$ & $0.42$ \\   \hline
 $  u = 5 $   & $0.67$  & $ 0.67 $ & $0.78$  & $ 0.78 $& $1.02$ & $1.03$ \\   \hline
 $  u = 10 $   & $0.75$  & $ 0.75 $ & $0.99$  & $ 1.00 $ & $1.12$ &  $1.13$ \\   \hline
 $  u = 20 $   & $0.96$  & $ 0.96 $ & $1.14$  & $1.14$ & $1.44$ & $1.46$ \\   \hline
 $  u = 30 $   & $1.14$  & $ 1.14$  & $1.59$  & $1.61$  & $1.73$ & $1.76$ \\   \hline
 $  u = 40 $   & $1.31$  & $ 1.31 $ &$1.78$  & $ 1.78$ & $2.13$ & $2.16 $\\   \hline
 $  u = 50 $   & $1.69$  & $ 1.68 $ & $1.95$  & $1.95$ & $2.46$ & $2.50$ \\   \hline
 $  u = 60 $   & $2.00$  & $1.98$  & $2.94$  & $2.93$ & $3.16$ & $3.18 $\\   \hline
 $  u = 80 $   & $2.87$ & $ 2.74$  & $5.20$ & $4.94$ & $5.72$ & $5.66$  \\   \hline
 $  u = 90 $   & $5.49$  & $ 4.30 $ & $8.27$ & $ 7.01 $&  $10.93$ & $9.53$ \\   \hline \hline
(B) & \multicolumn{2}{c |}{$r=150$} &  \multicolumn{2}{c |}{$r=250$} &  \multicolumn{2}{c |}{$r=350$} \\
Time  & $\semuhat $& $\spn(\Gammabfhat_{\oneD})$ & $\semuhat $& $\spn(\Gammabfhat_{\oneD})$ & $\semuhat $& $\spn(\Gammabfhat_{\oneD})$\\
   \hline\hline
 $  u = 1 $   & $0.16$  & $0.18 $& $0.45$ & $ 0.64$ &   $0.96$ & $1.65$ \\   \hline
 $  u = 5 $   & $0.21$  & $ 0.85 $ & $0.54$ & $ 3.30$& $1.08$ & $ 8.23 $\\  \hline
 $  u = 10 $   & $0.28$  & $ 2.26 $ & $0.64$  & $ 8.31$ & $1.22$ & $20.89 $\\   \hline
 $  u = 20 $   & $0.42$  & $6.16$ & $0.94$  & $23.4$ & $1.64$ & $51.61$ \\   \hline
 $  u = 30 $   & $0.62$ & $9.56$ & $1.33$ & $ 37.00$ &   $2.18$ & $81.46$\\   \hline
 $  u = 40 $   & $0.86$ & $ 12.94$ & $1.83$ & $ 48.45$ &   $2.86$ & $ 110.06$\\   \hline
 $  u = 50 $   & $1.09$  & $ 16.03$ & $2.38$ & $59.20$ &   $3.73$& $135.05$ \\   \hline
 $  u = 60 $   & $1.40$ & $ 18.62$ &  $3.13$ & $ 68.23$ &   $4.85$ & $ 157.65$ \\   \hline
 $  u = 80 $   & $2.08$  & $ 22.50$& $9.77$  & $87.08$ & $15.07$  & $196.84$ \\   \hline
 $  u = 90 $   & $2.50$ & $23.91$ & $11.74$  & $91.20$ & $27.97$ & $212.87$ \\   \hline

    \end{tabular}
\caption{Scenario VII.  (A) Angle between $\semu$ and the indicated subspace. (B)  Computing time in seconds for the indicated subspace.    $\semuhat$ and $\spn(\Gammabfhat_{\oneD})$ denote the subspaces by the new non Grassmann algorithm and the 1D algorithm. }
\label{tab:scenarioVII}
\end{table}

\section{Contrasts on real data} \label{sec:data}

In this section we compare the computing time for the new non Grassmann algorithm and the 1D algorithm to select an envelope dimension by minimizing prediction errors from five fold cross validation, the method typically used in conjunction with the 1D algorithm.  The time reported is, for each $u$,  the total optimization time over $250$ optimizations comprised of $50$ replications of five fold cross validation.  

\subsection{Alzheimer data}

\begin{figure}[ht]
\centering
\includegraphics[width=0.8\textwidth]{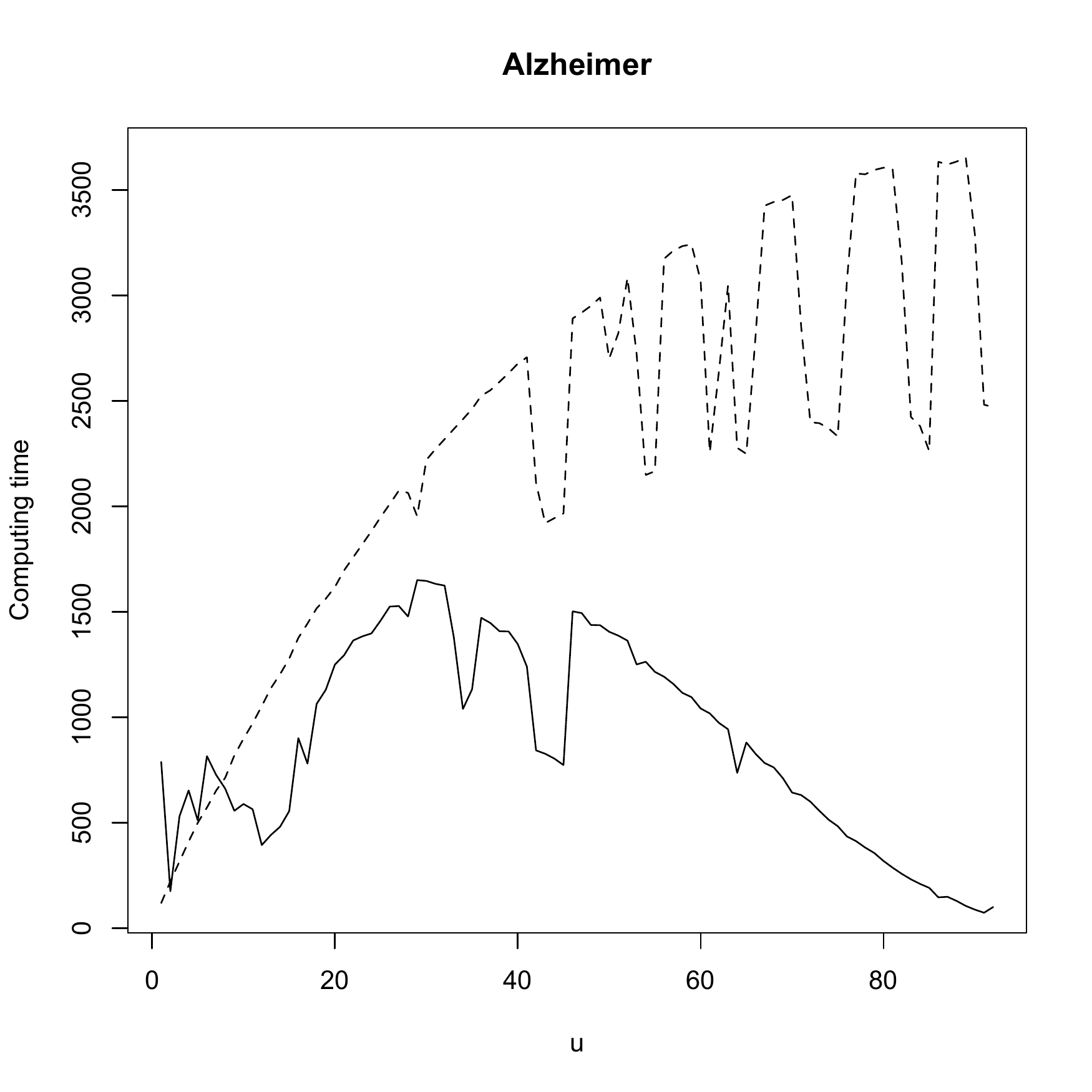}
\caption{Alzheimer data: for each $u$ the vertical axis is the total optimization time over $250$ optimizations comprised of $50$ replications of five fold cross validation. The solid line marks the new non Grassmann algorithm and the dashed line marks 1D algorithm.   }\label{malefemale:time}
\end{figure}

The Alzheimer data contains volumes of $r=93$ regions of the brain from each of  $749$ Alzheimer patients (Zhu et al. (2014)).   We used gender, age, the logarithm of intracerebroventricular volume, and interactions involving gender as predictors.  After taking the logarithms of all brain volumes, we fitted the response envelope model using both the new algorithm and the 1D algorithm.  There was little to distinguish methods based on predictive performance, but the time differences are clear, as displayed in Figure \ref{malefemale:time}.  As we observed in the simulations, the times for the two algorithms are close for relatively small values of $u$ and diverge for larger values of $u$.  The total optimization time over all $250 \times r = 23, 250$ optimizations was about 22 hours for the new algorithm and 60 hours for the 1D algorithm.  The overall computation time is relatively large because the signal in the data is somewhat weak.  

The value of $u$ selected by the new algorithm was  $u = 17$, while the 1D algorithm selected $u=6$.  The average prediction errors estimated by the two methods at their own values of $u$ were essentially the same.

\subsection{Glass data}

Our algorithm is also applicable in envelope contexts other than response envelopes.  We used predictor envelopes (Cook et al. (2013)) for this illustration.  

\begin{figure}[ht]
\centering
\includegraphics[width=0.8\textwidth]{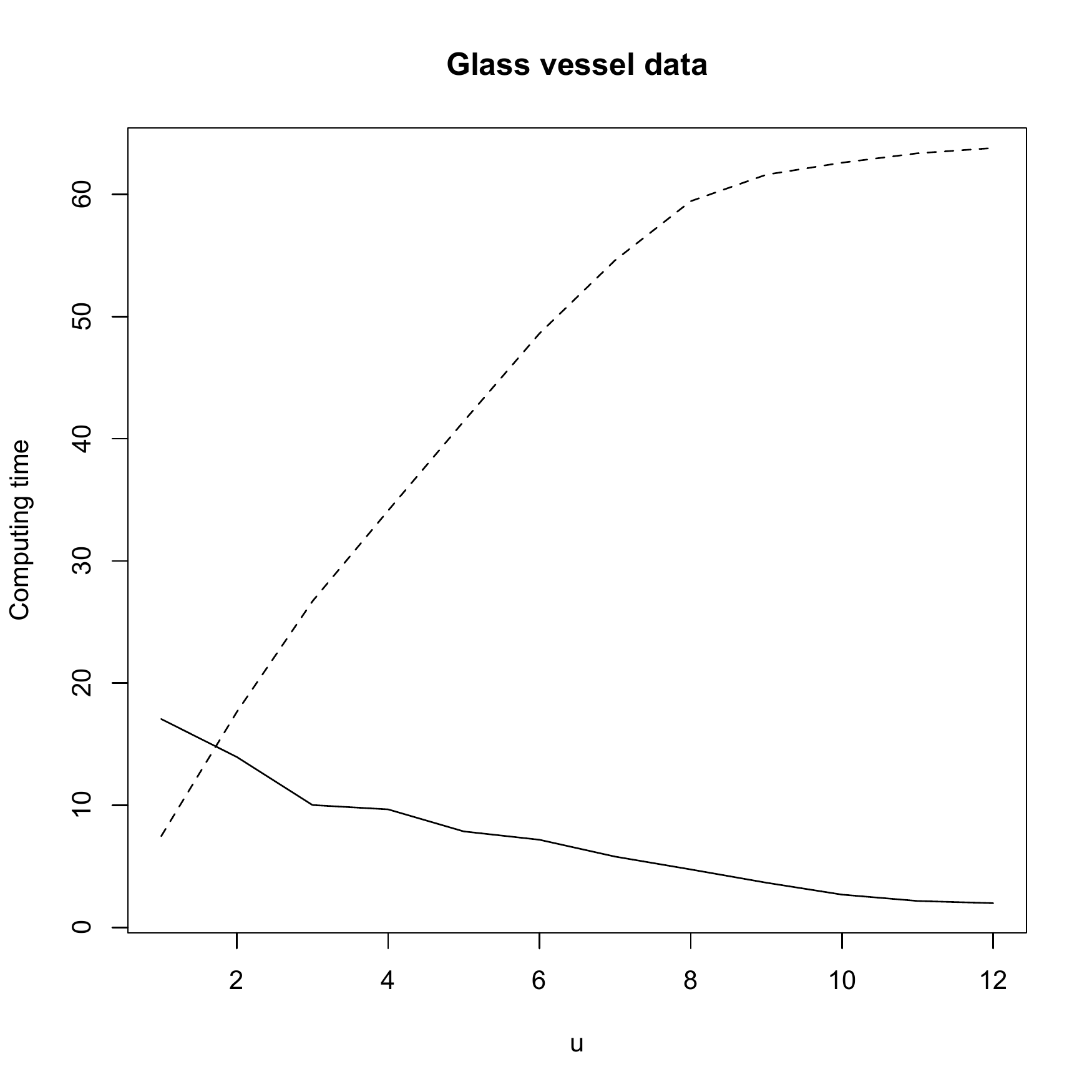}
\caption{The solid line marks the new non Grassmann algorithm and the dashed line marks 1D algorithm.  }\label{fig:glass-time}
\end{figure}

The dataset contains measurements of the chemical composition and electron-probe-X-ray microanalysis for $180$ archeological glass vessels from 15th to 17th century excavated in Antwerp, Belgium.  For each vessel, a spectrum on a set of equispaced frequencies between $1$ and $1920$ is measured, however the values below $100$ and above $400$ are almost null.  Following Kudraszow and Maronna (2011), we chose $13$ equispaced frequencies between $100$ and $400$ as predictors.  The response variable is the amount of sulfur trioxide. For each $u=1,\ldots,13$, we ran the 1D algorithm and the new algorithm, recoding the prediction error from 50 replications of five fold cross validation and the average computing time for these $250$ optimizations. The new algorithm gave a four percent improvement in prediction error over the 1D algorithm at  $u=3$, which was best for both methods.  As in the Alzheimer data, there were clear differences in computing time, as shown in Figure~\ref{fig:glass-time}.  The total time for computing all $u$ was $86$ seconds for the new algorithm and $541$ seconds for the 1D algorithm.

\section*{Acknowledgements}

The authors thank Hongtu Zhu for providing the Alzheimer data and Nadia L. Kudraszow for providing the glass data.   Research for this article was supported in part by grant DMS-1407460 from the National Science Foundation.

  \newpage{}

\appendix
 \global\long\def\thesection{\Alph{section}}
 \setcounter{equation}{0} 
\begin{appendix} 
\begin{center}\LARGE{\bf Appendix}
\end{center}
\section{Proof of Proposition \ref{prop:min1}}

Let $ (\G, \G_{0}) \in \real{r \times r}$ be a column partitioned orthogonal matrix and let $\M \in \sreal{r \times r}$ be positive definite.  The conclusion that $\log|\G^{T}\M\G| + \log|\G_{0}\M\G_{0}|$ is minimized when $\spn(\G)$ is any $u$-dimensional reducing subspace of $\M$ will follow by showing that $|\M| \leq |\G^{T}\M\G| \times |\G_{0}^{T}\M\G_{0}|$ with equality if and only if $\spn(\G)$ reduces $\M$.
\begin{eqnarray*}
|\M| &  = &  |(\G,\G_{0})^{T}\M(\G,\G_{0})| = \left| \begin{array}{cc} \G^{T}\M\G & \G^{T}\M\G_{0} \\ \G_{0}^{T}\M\G & \G_{0}^{T}\M\G_{0} \end{array} \right| \\
& = &| \G^{T}\M\G | \times |\G_{0}^{T}\M\G_{0} - \G_{0}^{T}\M\G (\G^{T}\M\G )^{-1}\G^{T}\M\G_{0} | \\
& \leq & | \G^{T}\M\G_{1} | \times |\G_{0}^{T}\M\G_{0}|,
\end{eqnarray*}
with equality if and only if $ \G_{0}^{T}\M\G = 0$, which is equivalent to requiring that $\spn(\G)$ reduce $\M$.  

The conclusion that $\log|\G^{T}\M\G| + \log|\G\M^{-1}\G|$ is also minimized when $\spn(\G)$ is any $u$-dimensional reducing subspace of $\M$  follows because 
\[
\log|\G^{T}\M\G| + \log|\G\M^{-1}\G| = \log|\G^{T}\M\G| + \log|\G_{0}\M\G_{0}| - \log|\M|.
\]
\section{Proof of Proposition \ref{prop:min2}}

Recall that  $J_{1}(\G) =  \log|\G^T\Mhat\G| + \log |\G_{0}^T\Mhat\G_{0}| $, $J_{2}(\G) = \log | \I_{r-u} +\G_{0}^{T}\Uhat_{\M}\G_{0}|$ and $J(\G) = J_{1}(\G) + J_{2}(\G)$, where $\Uhat_{\M} = \Mhat^{-1/2}\Uhat\Mhat^{-1/2}$ is a standardized version of $\Uhat$.  Then
from Proposition~\ref{prop:min1}, an argument  minimizes $L_{u}(\G)$ if and only if it minimizes
\begin{eqnarray*}
f(\G) & = & \log|\G^T\Mhat\G| + \log|\G_{0}^T(\Mhat +\Uhat)\G_{0}| \\
& = & \log|\G^T\Mhat\G| + \log|\G_{0}^T(\Mhat +\ubfhat\ubfhat^{T})\G_{0}|\\
& = &  \log|\G^T\Mhat\G| + \log |\G_{0}^T\Mhat\G_{0}| + \log | \I_{k} + \ubfhat^{T}\G_{0}(\G_{0}^{T}\Mhat\G_{0})^{-1}\G_{0}^{T}\ubfhat | \\
& = &   \log|\G^T\Mhat\G| + \log |\G_{0}^T\Mhat\G_{0}| \\
& & + \log | \I_{r-u} +(\G_{0}^{T}\Mhat\G_{0})^{-1/2}\G_{0}^{T}\ubfhat \ubfhat^{T}\G_{0}(\G_{0}^{T}\Mhat\G_{0})^{-1/2}| \\
& = & J_{1}(\G) + f_{2}(\G),
\end{eqnarray*}
where $f_{2}$ is defined implicitly and $\Uhat = \ubfhat \ubfhat^{T}$ is a decomposition of $\Uhat$ with $\ubfhat \in \real{r \times k}$.  To see that $f_{2} = J_{2}$  over $\vspc_u$ we have
\begin{eqnarray*}
f_{2}(\G) & = & \log | \I_{r-u} +(\G_{0}^{T}\Mhat\G_{0})^{-1/2}\G_{0}^{T}\Uhat\G_{0}(\G_{0}^{T}\Mhat\G_{0})^{-1/2}| \\
& = &  \log | \I_{r-u} +(\G_{0}^{T}\Mhat\G_{0})^{-1/2}\G_{0}^{T}\Mhat^{1/2}(\Mhat^{-1/2}\Uhat\Mhat^{-1/2})\Mhat^{1/2}\G_{0}(\G_{0}^{T}\Mhat\G_{0})^{-1/2}| \\
& = &  \log | \I_{r-u} +\G_{0}^{T}(\Mhat^{-1/2}\Uhat\Mhat^{-1/2})\G_{0}| \\
& = & \log | \I_{r-u} +\G_{0}^{T}\Uhat_{\M}\G_{0}| \\
& = & J_{2}(\G),
\end{eqnarray*}
where the third equality follows because $\G_{0} \in \vspc_{u}$ reduces $\Mhat$.

\section{Proof of Proposition \ref{prop:min3}}

Let $\What = \Mhat + \Uhat$ for notational convenience and start with the objective function
\begin{eqnarray*}
L_{u}(\G) &  = & \log|\G^T\Mhat\G| + \log|\G^T(\Mhat +\Uhat)^{-1}\G| \\
&  = & \log|\G^T\Mhat\G| + \log|\G^T\What^{-1}\G| \\
& = &  \log|\G^T(\Mhat+ \Uhat)\G - \G^{T}\Uhat\G| + \log|\G^T\What^{-1}\G| \\
& = &  \log|\G^T\What\G - \G^{T}\ubfhat\ubfhat^{T}\G| + \log|\G^T\What^{-1}\G| \\
& = &  \log|\G^T\What\G| +  \log|\I_{k} - \ubfhat^{T}\G(\G^T\What\G)^{-1}\G^{T}\ubfhat| + \log|\G^T\What^{-1}\G|,
\end{eqnarray*}
where $\ubfhat$ is as defined in the proof of Proposition~\ref{prop:min2}.
The sum of the first and last terms on the right side of this representation is always non-negative and equals 0, its minimum value, when the columns of $\G$ span any reducing subspace of $\What = \Mhat + \Uhat$.  Restricting $\G$ in this way,
\begin{eqnarray*}
\ubfhat^{T}\G(\G^T\What\G)^{-1}\G^{T}\ubfhat & = & \ubfhat^{T}\What^{-1/2}\What^{1/2}\G(\G^T\What\G)^{-1}\G^{T}\What^{1/2}\What^{-1/2}\ubfhat \\
& = &  \ubfhat^{T}\What^{-1/2}\G\G^{T}\What^{-1/2}\ubfhat,
\end{eqnarray*}
and the middle term of $L(\G)$ reduces to
\begin{eqnarray*}
 \log|\I - \ubfhat^{T}\G(\G^T(\Mhat+ \Uhat)\G)^{-1}\G^{T}\ubfhat| & =  & \log|\I_{k} - \ubfhat^{T}\What^{-1/2}\G\G^{T}\What^{-1/2}\ubfhat| \\
  & = &   \log|\I_{u}- \G^{T}\What^{-1/2}\ubfhat\ubfhat^{T}\What^{-1/2}\G| \\
  & = &  \log|\I_{u} - \G^{T}\Uhat_{\M+\U}\G|,
  \end{eqnarray*}
  where $\Uhat_{\M+\U} = \What^{-1/2}\ubfhat\ubfhat^{T}\What^{-1/2} = \What^{-1/2}\Uhat\What^{-1/2}$ is $\Uhat$ standardized by $\What^{-1/2} = (\Mhat + \Uhat)^{-1/2}$. 
  
\section{Proof of Proposition \ref{prop:consistency}}
We demonstrate the result in detail for $K_{\M}$.  The corresponding result for the other three $K$ functions follows similarly.

Recall that $K_{\M}(\G) = \sum_{i=1}^{u} (\g_{i}^{T}\Mhat^{-1/2}\Uhat\Mhat^{-1/2}\g_{i})$ where $\g_{i}$ is an eigenvector of $\Mhat$.  
The population version of this objective function is 
\[
\tilde{K}_{M}(\tilde{\G})= \sum_{i=1}^{u} (\gt_{i}^{T}\M^{-1/2}\U\M^{-1/2}\gt_{i}) 
\]
where $\gt$ is an eigenvector of $\M$ and $\tilde{\G} = (\gt_{1},\ldots,\gt_{u})$.  We next show that 
\[
\spn\left(\arg \max \tilde{K}_{\M}(\tilde{\G})\right) = \semu.
\]

Consider a generic envelope $\espc_{\A}(\spc)$, where $\A > 0$ with eigenspaces $\aspc_{i}$, $i=1,\ldots,q$.  Cook, et al. (2010) show that this envelope can be characterizes as  $\espc_{\A}(\spc) = \sum_{i=1}^{q} \Pbf_{\aspc_{i}} \spc$.  As a consequence there are $u = \dim(\espc_{\A}(\spc))$ orthogonal eigenvectors $\abf_{1},\ldots \abf_{u}$ of $\A$ so 
that  
\[
\espc_{\A}(\spc) = \sum_{i=1}^{u} \Pbf_{\abf_{i}} \spc = \spn\left(\sum_{i=1}^{u} \Pbf_{\abf_{i}} \s\s^{T} \Pbf_{\abf_{i}}\right)
\]
where $\s$ is a basis matrix for $\spc$.  By definition of $\espc_{\A}(\spc)$, there exists exactly $u$ eigenvectors of $\A$ that are not orthogonal to $\spc$ and these eigenvectors are $\abf_{1},\ldots \abf_{u}$.  Consequently, we must have
\[
\espc_{\A}(\spc) = \spn \left(\arg \max \tr\left(\sum_{i=1}^{u} \Pbf_{\bv_{i}} \s\s^{T} \Pbf_{\bv_{i}}\right)\right) = \spn \left(\arg \max \sum_{i=1}^{u} \bv_{i}^{T} \s\s^{T} \bv_{i}\right),
\]
where the maximum is taken over the eigenvectors $\bv_{i}$ of $\A$.  Equality holds since the maximum must select $u$ eigenvectors of $\A$ that are not orthogonal to $\s\s^{T}$.

Comparing this general argument with $\tilde{K}(\G)$ we see that $\arg \max \tilde{K}$ will select $u$ eigenvectors of $\M$ that are not orthogonal to $\M^{-1/2}\U\M^{-1/2}$ and consequently
\[
\spn\left(\arg \max \tilde{K}_{\M}(\tilde{\G})\right) = \espc_{\M}(\spn(\M^{-1/2}\U\M^{-1/2})) = \semu,
\]
where the final equality follows from Cook et al. (2010, Prop. 2.4).

The $\sqrt{n}$ consistency now follows straightforwardly since the matrices involved in the determination of the four potential starting values -- $\Mhat$, $\Mhat + \Uhat$, $\Uhat_{\M}$ and $\Uhat_{\M + \U}$ -- are all $\sqrt{n}$-consistent estimators of their corresponding population versions.

 \section{Proof of Proposition \ref{prop:ge}}
Let $\Gammabf_{\start}\in\real{r\times u}$ denote the population counterpart of $\G_{\start}$.  Based on previous discussion, the columns of $\Gammabf_{\start}$ are eigenvectors of $\M$ or $\M+\U$.  
Since $\hbox{rank}( \Gammabf_{\start} )=u$ we can find $u$ linearly independent rows of $\Gammabf_\start$  
 and, letting $\Gammabf_{ u}$ denote the $u\times u$ matrix forms by these $u$ rows, we get $|\Gammabf_{ u} |\not=0$.
Now, let $ \G_{ u} $ denote the submatrix of ${\G}_{\start}$ forms by these same $u$ rows. It follows straightforwardly in the manner of Proposition~\ref{prop:consistency} that $ \G_{u} $ is a $\sqrt{n}$ consistent estimator of $\Gammabf_u$. Since the determinant is a continuous function this implies that for $n$ sufficiently large $| \G_{u} |\not=0$ with a specified high probability. As a consequence, for $n$ sufficiently large,  $\hbox{rank} (\G_{\start}) =u$ with arbitrarily high probability.

Perform Gaussian elimination with partial pivoting on $\G_{\start}$ and denote the resulting $u\times u$ submatrix by $\G_{\start, u}$.  From the preceding discussion,    $\G_{\start, u}$ is nonsingular with high probability for sufficiently large $n$.
Also, perform Gaussian elimination with partial pivoting to $\Gammabf_{\start}$   and denote the resulting nonsingular $u\times u$ submatrix by $\Gammabf_{\start, u}$.  The proposition is then established if $\G_{\start, u}$ is a $\sqrt{n}$ consistent estimator of $\Gammabf_{\start, u}$.  

%
First we assume that the pivot elements for $\Gammabf_{\start}$ are unique and occur in rows $r_{i}$, $i=1, \ldots, u$. 
In the first step of Gaussian elimination, for an arbitrary $\epsilon>0$, we can find an $N_{1}$ such that when $n>N_{1}$, the corresponding element in row $r_{1}$ of $\G_{\start}$ is the one having the largest absolute value with probability at least $1-\epsilon$. In other words, row $r_{1}$ will be selected in $\G_{\start}$ with probability at least $1-\epsilon$.  We call the resulting matrices
$\Gammabf_{\start, 1}\in\real{r\times u}$ and $\G_{\start, 1}\in\real{r\times u} $. 
As Gaussian elimination  involves only simple arithmetic operations, $\G_{\start, 1}$ converges to $\Gammabf_{\start, 1}$ at rate $\sqrt{n}$.
Now, for the second step in Gaussian elimination, we do partial pivoting in the second columns of  $\G_{\start,1}$  and $\Gammabf_{\start,1}$.
Then, for an arbitrary $\epsilon>0$, we can find an $N_{2}>N_{1}$ such that when $n>N_{2}$, the  elements chosen for $\G_{\start,1}$ and $\Gammabf_{\start,1}$ will be the same with probability at least $(1-\epsilon)$.

Continuing this process, for $n> N_{u}$, rows $r_{1}, \ldots, r_{u}$ in $\G_{\start}$ are selected with probability at least $(1-\epsilon)^{u}$.  Let $\| \cdot\|$ denote some matrix norm.  As $\G_{\start}$ converges to $\Gammabf_{\start}$ with rate $\sqrt{n}$, we have $\|\G_{\start, u} - \Gammabf_{\start, u}\| = O_{p}(n^{-1/2})$ and consequently for any $\epsilon > 0$ there exists $K>0$ and $N_{0}$ so that for all $n>N_{0}$, 
\[
P\left( \sqrt{n}\|\G_{\start, u}-\Gammabf_{\start, u}\|  > K \;\bigg| \; \text{rows}\; r_{1}, \ldots r_{u} \;\text{are selected}\right) < \epsilon.
\]
Then with $n>\max(N_{0}, N_{u})$,
\begin{eqnarray*}
& &P\left( \sqrt{n}\|\G_{\start, u}-\Gammabf_{\start, u}\| > K\right)\\
& < & P\left(\sqrt{n}\|\G_{\start, u}-\Gammabf_{\start, u}\|  > K \; \bigg| \; \text{rows}\; r_{1}, \ldots r_{u} \;\text{are selected}\right) * P(\text{rows}\; r_{1}, \ldots r_{u} \;\text{are selected})\\
& & + P(\text{not all rows}\; r_{1}, \ldots r_{u}\; \text{are selected})\\
 &<& \epsilon  + [1 - (1-\epsilon)^{u}]. 
\end{eqnarray*}
Since $\epsilon > 0$ is arbitrary and $\epsilon + [1 - (1-\epsilon)^{u}]$ tends to $0$ as $\epsilon$ tends to $0$, $\G_{\start, u}$ converges to $\Gammabf_{\start, u}$ at rate $\sqrt{n}$.

To deal with non-unique pivot elements, assume that there are ties in one column.
When we perform Gaussian elimination with partial pivoting   on $\Gammabf_\start$ in the step  with $k$ ties, 
we can choose whichever of the tied elements, resulting in all the cases in  non-singular matrices.  We call the resulting matrices $\A_{1} , \dots, \A_{k}$. 
When  Gaussian elimination was perform with partial pivoting on $\G_{\start}$, using the preceding reasoning, there will be probability at least $(1-\epsilon)^{u}/k$ that we pick the rows in $\A_{i}$, $i=1,\dots,k$.  Then $\Ahat$ converges to $\A_{1}, \A_{2} \dots $ or $\A_{k}$ with rate $\sqrt{n}$, so $\Ahat$ converges to a non-singular matrix with rate $\sqrt{n}$.
If we have ties in more than one step we divide further probabilities, since the number of the steps and $u$ are fixed  the proof flows similarly.  

\end{appendix}


\begin{thebibliography}{References}

\bibitem{CHS} Cook, R.D., Helland, I. S. and Su, Z. (2013), Envelopes and partial least squares regression. {\em Journal of the Royal Statistical Society B} {\bf 75}, 851--877.

\bibitem{CLC07} Cook, R.D., Li, B. and Chiaromonte, F. (2007).  Dimension reduction in regression without matrix inversion.  {\em Biometrika} {\bf 94}, 569--584.

\bibitem{CLC} Cook, R.D., Li, B. and Chiaromonte, F. (2010). Envelope models for parsimonious and efficient multivariate linear
regression (with discussion). {\em Statistica Sinica}, \textbf{20}, 927--1010.


\bibitem{CSY} Cook, R. D., Su, Z. and Yang, Y. (2014).  A MATLAB toolbox for computing envelope estimators in multivariate analyses.  {\em Journal of Statistical Software} {\bf 62}, http://www.jstatsoft.org/v62/i08/paper.

\bibitem{CZfound} Cook, R.D. and Zhang, X.  (2015a). Foundations for envelope models and methods. {\em Journal of the American Statistical Association} {\bf 110}, 599--611.

\bibitem{CZsim} Cook, R.D. and Zhang, X.  (2015b). Simultaneous envelopes for multivariate linear regression. {\em Technometrics} {\bf 57}, 11--25.

\bibitem{CZalg} Cook, R.D. and Zhang, X.  (2015c). Algorithms for envelope estimation. {\em Journal of Computational and Graphical Statistics}, to appear (http://arxiv.org/pdf/1403.4138.pdf)

\bibitem{Edelman} Edelman, A.,  Arias, T. A. and Smith, S. T. (1998). The geometry of algorithms with orthogonality constraints.
{\em SIAM Journal on Matrix Analysis and Applications}
{\bf 20}, 303 -- 353.


\bibitem{gv}  Kudraszow, N. L. and Maronna, R. A. (2011). Estimates of MM type for the multivariate linear model. {\it Journal of Multivariate Analysis}, {\bf 102} 1280--1292.

\bibitem{sc} Su, Z. and Cook, R.D.  (2011), Partial envelopes
for efficient estimation in multivariate linear regression. {\em
Biometrika}, \textbf{98}, 133--146.

\bibitem{SChetero} Su, Z. and Cook, R.D.  (2013). Estimation of multivariate means with heteroscedastic errors using envelope models.  {\em
Statistica Sinica}, \textbf{23}, 213--230.

\bibitem{azhm} Zhu, H., Khondker, Z., Lu, Z., and Ibrahim, J. G. (2014). Bayesian Generalized Low Rank Regression Models for Neuroimaging Phenotypes and Genetic Markers. {\it Journal of the American Statistical Association}, {\bf 109}, 977--990.
  \end{thebibliography}
\end{document}